\documentclass[12pt,preprint]{aastex}

\input{epsf}
\def\n{{\noindent}}
\newcommand{\be}{\begin{equation}}
\newcommand{\ee}{\end{equation}}
\newcommand{\ba}{\begin{eqnarray}}
\newcommand{\ea}{\end{eqnarray}}

\shorttitle{Weak Lensing and Dark Energy}
\shortauthors{Munshi \& Wang}

\begin{document}

\title{How Sensitive Are Weak Lensing Statistics \\
to Dark Energy Content?}
\author{Dipak Munshi$^{1,2}$, and Yun Wang$^{3}$}
\affil{{}$^1$Institute of Astronomy, Madingley Road,
Cambridge, CB3 OHA, United Kingdom\\
{}$^2$Astrophysics Group, Cavendish Laboratory, Madingley Road, Cambridge,
CB3 OHE, United Kingdom\\
{}$^3$Department of Physics \& Astronomy,University of Oklahoma, 
Norman, OK 73019 USA. munshi@ast.cam.ac.uk, wang@mail.nhn.ou.edu}

\begin{abstract}
Future weak lensing surveys will directly probe the clustering of dark
matter, in addition to providing a test for various cosmological models.
Recent studies have provided us with the tools
which can be used to construct the complete probability distribution
function for convergence fields. It is also possible to construct
the bias associated with the hot-spots in convergence maps. These
techniques can be used in both the quasi-linear and the highly nonlinear 
regimes using various well developed numerical methods. We use these results
here to study the weak lensing statistics of cosmological models
with dark energy. We study how well various classes of 
dark energy models can be distinguished from models with a cosmological
constant. We find that the ratio of the square root of the variance of
convergence is complementary to the convergence skewness $S_3$ in probing 
dark energy equation of state; it can be used to predict the expected
difference in weak lensing statistics between various dark energy models,
and for choosing optimized smoothing angles to constrain a given class of
dark energy models. 
Our results should be useful for probing dark energy
using future weak lensing data with high statistics from galaxy weak 
lensing surveys and supernova pencil beam surveys.
\end {abstract}




\section{Introduction}

Recent cosmological observations favor an accelerating universe
\cite{Garna98a,Riess98,Perl99}.
This implies the existence of energy of unknown nature (dark energy), 
which has negative pressure. Current data are consistent with dark energy
being a non-zero cosmological constant (see for example, Wang \& Garnavich 2001;
Bean \& Melchiorri 2002). Many other alternative
dark energy candidates have been considered, and are consistent with
data as well. For example, quintessence, k-essence, spintessence, etc.
(Peebles \& Ratra 1988; Frieman et al. 1995; Caldwell, Dave, \& Steinhardt 1998;
Garnavich et al. 1998b; White 1998; Efstathiou 1999; Steinhardt, Wang, \& Zlatev 1999;
Podariu \& Ratra 2000; Sahni \& Wang 2000; Sahni \& Starobinsky 2000; 
Saini et al. 2000; Waga \& Frieman 2000;
Huterer \& Turner 2001; Ng \& Wiltshire 2001; Podariu, Nugent, \& Ratra 2001;
Weller \& Albrecht 2001)

Different dark energy models can be conveniently classified according
to the equation of state of the dark energy component, $w_X$. For
example, for quintessence models, $dw_X /dz > 0$, while for k-essence
models, $dw_X /dz < 0$. There are many complimentary probes of dark energy.
These include, the distance-redshift relations of cosmological standard
candles, Cosmic Microwave Background Anisotropy, volume-redshift test
using galaxy counts, the evolution of galaxy clustering, weak lensing,
etc. These different methods to probe dark energy are complimentary
to each other, and can provide important consistency checks,
due to the different sources of systematics in each method
(for example, see Kujat et al. 2002).

Weak lensing surveys (Bacon, Refregier \& Ellis 2000; 
Van Waerbeke et al. 2000; Wittman et al. 2000;
Maoli et al. 2001; Van Waerbeke et al. 2001;
Wilson, Kaiser, \& Luppino 2001; Bacon et al. 2002; Hoekstra et al. 2002; 
Refregier, Rhodes \& Groth 2002), currently underway and more 
proposed in the near future, are well suited to studying the dark energy 
equation of state.
Weak lensing directly probes the gravitational clustering
and the background cosmology. Many recent studies, both theoretical and
numerical, have analyzed these possibilities. Observational teams have already
reported first detections of cosmic shear. On theoretical front  
progress has been made in modeling the statistics using both 
perturbative calculations which are valid for large smoothing angles and
also using the well motivated hierarchical ansatz which is valid for
small smoothing angles. Numerical studies carried out so far uses ray shooting
experiments and are quite useful in testing analytical calculations.

In an earlier study on probing quintessence using weak lensing, 
Hui (1999) concluded that the large scale 
convergence skewness can directly provide a constraint
for $w_X$, the equation of state of dark energy. 
Similarly, Huterer studied the use of weak lensing convergence 
power spectrum and three-point statistics to constrain dark energy models.
Both of these papers are useful in utilizing large weak lensing surveys of 
galaxies to constrain dark energy. 

In this paper we construct the complete probability distribution function
of convergence
to study the effects of dark energy. Our results apply to the weak lensing
of both galaxies (on large angular scales) and type Ia supernovae (on
small angular scales). We compare two classes of dark energy models
(one with effective constant equation of state $w_X$, the other with
time-varying $w_X$) against that of a $\Lambda$ dominated universe.  

The technique we use in this paper has been tested in detail using
N-body calculations. Analytical results
are obtained for large smoothing angles using the perturbative calculations,
and for small smoothing angles using the hierarchical ansatz.
We focus on both the one-point probability distribution function and 
the bias associated with convergence maps in the quasi-linear 
and the highly nonlinear regimes. Our studies are quite complementary 
to the studies done using a Fisher matrix analysis for the recovery
of power spectra from observations. While such studies are well suited for
recovering the nonlinear matter power spectra; the study of the probability
distribution function and the bias will give us a direct probe
of non-Gaussianity developed through gravitational clustering.

The paper is organized as follows. In Section $2$ we give the definition
of some basic equations for reference, and identify the dark
energy models studied in this paper.
Section $3$ discusses the weak lensing statistics of dark energy models.
Section $4$ contains discussions and a summary.

\section{Notation}

\n
The  weak lensing convergence, $\kappa$, maps the distribution of projected 
density fields, and its statistics is directly related to that of the
underlying matter distribution. We write

\begin{equation}
\kappa(\theta_0) = \int_0^{\chi_s} d \chi \omega(\chi) 
\delta(r(\chi)\theta_0,\chi),
\end{equation}

\n
where $r(\chi)$ is the angular diameter distance and $\omega(\chi)$ is the 
weight function associated with the source distribution. The observer is 
placed at $\chi=0$, and the sources (which for simplicity we assume are all 
at the same redshift) are placed at $\chi_s$. $\chi$ is given by 

\begin{equation}
\chi(z) = cH_0^{-1} \int_0^z dz'[\Omega_m(1+z')^3 + \Omega_k(1+z')^2 + 
\Omega_X \, f(z) ]^{-1/2}
\end{equation}

\n
where the $\Omega$'s denote the fraction of critical density in various 
components.  $\Omega_X$ denotes the dark energy component,
and $\Omega_k=1-\Omega_m-\Omega_X$, The function
$f(z)$ parametrizes the time-dependence of the dark energy density,
and $f(z=0)=1$. For dark energy with constant equation of state, 
$w_X= p_X/\rho_X = constant$, $f(z)=(1 + z)^{3(1+w_X)}$.
The limiting case with $w_X = -1$ [i.e.,
$f(z)=1$] corresponds to a cosmological constant. 
Note that to obtain accelerated expansion of the universe,
we need $\rho+3 p <0$, which implies $w_X <-1/3$ for a power law
dark energy density $f(z)$. 
In general, the dark energy equation of state, $w_X(z)$,
can be written in terms of the dimensionless dark energy density, $f(z)$,
$w_X(z)=\frac{1}{3}(1+z)\,f'(z)/f(z) -1$, for dark energy density or 
equation of state with arbitrary time dependence.
(Wang \& Garnavich 2001)

We study two classes of dark energy models. 
The first class contains dark energy models with effective constant 
equation of state, $w_X=-1/3, -2/3, -1$ ($\Lambda$CDM), and $-1.9$.
If dark energy arises from classical fields, it must satisfy the
weak energy condition, which requires that $w_X >-1$.
The weak energy condition violating toy model, $w_X=-1.9$,
is motivated by quantum gravity models of inflation in which 
quantum effects lead to the violation of the weak energy condition
(Onemli \& Woodard 2002).
The second class contains two dark energy toy models
with linear time-varying equations of state, $w_X=w_q(z)=-1+z$ and $-1+2z/3$. 
These are motivated by quintessence models which have $w_X$ that effectively
increases with $z$.
All the models have $\Omega_m=0.3$, $\Omega_X=0.7$, $h=0.7$, $n_S=1$ 
(power law index of the primordial power spectrum),
and $\sigma_8=0.8$. We normalize the nonlinear power spectrum
to $\sigma_8$. Table 1 lists these dark energy models.

\begin{center}
Table 1\\
{\footnotesize{Dark Energy Models}}

{\footnotesize
\begin{tabular}{|llll|}
\hline 
Model && $w_X(z)$   \\  
\hline 
$\Lambda$CDM 	&  $-1$       &        & \\
constant $w_X$  &  $-1/3$,    &  $-2/3$, & $-1.9$\\
$w_q(z)$ 		&  $-1+2z/3$, & $-1+z$   &    \\
\hline 
\end{tabular}
}
\end{center}

\n
The two classes of dark energy models are compared with the fiducial 
$\Lambda$CDM model to quantify the variation in various statistics of 
convergence maps.

\section{Statistics of Weak Lensing in Dark Energy Models} 

To compute the statistics of weak lensing convergence field, we 
need to relate it to the statistics of three dimensional density
field of underlying matter distribution. In recent studies such 
analysis has been done extensively. We use such a formalism to 
explore the weak lensing statistics for the dark energy cosmologies.
For large smoothing angles we use the perturbative calculations and
for small smoothing angles we use the well motivated hierarchical
ansatz to compute the relevant quantities.

\subsection{Evolution of the Matter Power Spectrum}

We compute the matter power spectrum using the scaling
ansatz of Hamilton et al. (1991),  which was later extended by various
authors [see e.g. Peacock \& Dodds (1996)]. This ansatz essentially
consists of postulating that $4\pi k^3 P(k) = f[ 4 \pi k_l^3 P_l(k_l)]$,
where $P(k)$ is the nonlinear power spectrum and $P_l$ is the linear
power spectrum, and the function $f$ in general will depend on the
initial power spectrum. The linear power spectrum is evaluated at a
different wave number, $k_l = ( 1 + 4 \pi k^3 P(k) )^{-1/3} k$, 
hence the mapping is non-local in nature. The cosmological model enters 
through the linear growth function $g(z)$, so that 
$P_l(k, z) = [ g(z)/(1+z)]^2\, P_l(k, z=0)$.
The linear growth function is given by:

\begin{equation}
g(z) = \frac{\delta(z, \Omega_m, \Omega_{\Lambda})}{\delta(z, \Omega_m=1)}
= { 5 \over 2 } \Omega_m \, (1+z)\, E(z)\, \int_z^{\infty} dz' \,
\frac{ (1+z')}{ \left[ E(z') \right]^3},
\end{equation} 

\n
Where 

\begin{equation}
E(z)  \equiv  \sqrt{ \Omega_m(1+z)^3 + \Omega_k(1+z)^2 + \Omega_X \, f(z)}
\end{equation}

\n
We compute the linear growth function by direct integration [for
more on power spectrum evolution in quintessence models see \cite{BeBa}].

Our method enforces stable clustering in the nonlinear regime and assumes the 
hierarchical ansatz (which is tested by numerous numerical simulations),
therefore we are able to predict the higher order correlation functions
(Davis \& Peebles 1977, Groth \& Peebles 1977, Fry \& Peebles 1978) and 
(their Fourier transforms or)  the multi-spectrum correctly. Combining these 
with the powerful technique of the generating function we can construct the 
complete probability distribution function and the bias associated with 
convergence maps.

\subsection{Convergence Probability Distribution Function}

Perturbative calculations depend on the expansion of the convergence
field $\kappa(\theta_0)$ for smoothing angle $\theta_0$ in terms 
of perturbative expansion of the density field $\delta$. Such an
analysis can be performed in an order by order manner,

\begin{equation}
\kappa^{(1)}(\theta_0) + \kappa^{(2)}(\theta_0) + \dots = 
\int_0^{\chi_s} d \chi \omega(\chi) \,
\left[\delta^{(1)}(r(\chi)\theta_0) +\delta^{(1)}(r(\chi)\theta_0)+ \dots \right] 
\end{equation}

\n
where $\delta^{(i)}$ and $\kappa^{(i)}$ correspond to the $i$-th order perturbative
expansion, $i=1$ being the linear order. In the perturbative regime at tree 
level (Fry 1984; Bernardeau 1992, 1994; Bernardeau \& Schaeffer 1992),
it is possible to introduce vertex generating function $G(\tau)$ which 
will encode the tree level contribution from all orders. 
The smoothing using a top-hat filter function can be incorporated
in the generating function formalism and then the generating function
can be written in terms of the generating function of the
unsmoothed case.  
All statistical quantities including probability distribution functions 
can be constructed once we have solved for the tree-level generating
functions (Bernardeau 1992, 1994). We write

\begin{equation}
G(\tau) = \left ( 1 - { \tau \over \kappa_a} \right )^{-\kappa_a};~~~
G_s^{PT}(\tau) = G^{PT} \left [ \tau { \sigma(R_0(1+G^{PT}(\tau))^{1/2})
\over \sigma(R_0) } \right ]
\end{equation}

\n
The parameter $\kappa_a$ can be determined from the dynamical equation 
governing the evolution of perturbations in the quasi-linear regime,
and it is given by $ \kappa_a = {{\sqrt {13} - 1} \over 2}$.
The variance at a length scale $\sigma(R_0) = R_0^{-(n+3)/2}$, where
a local power law spectrum index $n$ is used to evaluate the generating 
function. The generating function is now used to compute the probability 
distribution function at a particular smoothing scale.

In the highly nonlinear regime (Balian \& Schaeffer 1989, 
Davis \& Peebles 1977, Groth \& Peebels 1977, Szapudi \& Szalay 1993, 
Scoccimarro \& Frieman 1999, Munshi et al. 1999c), the perturbative series 
starts to diverge and the usual perturbative calculations are replaced by 
the hierarchical ansatz for higher order correlation functions, which 
can be built from two-point correlation functions. The amplitude
of various contributions can be constructed from the knowledge of the 
generating function. It was found from analytical reasoning and numerical 
experimentation that the generating function in the highly nonlinear regime
retains exactly the same form as in the quasi-linear regime; however, the 
value of $\kappa_a$ is changed (Beranrdeau 1992) -- it is now treated as 
a free parameter (Colombi et al. 1995, Munshi et al. 1999a,b). It is 
customary to use a different parameter $\omega$ that is easy to evaluate 
from numerical simulations,  $\kappa_a = {2 \omega \over (1 - \omega)}$. 
It was found that $\omega=.3$ reproduces various statistics in the
nonlinear regime quite well (Colombi et al. 1997, Colombi, Bouchet, 
Schaeffer 1995, Munshi et al. 1999d).  To compute the probability distribution
function, one has to compute the void probability function $\phi(y)$, 
which acts as a generating function for normalized cumulants or 
$S_N$ parameters and can be expressed in terms of the function $G(\tau)$ as
(Balian \& Schaeffer 1989):
\begin{eqnarray}
&&\phi(y) = y G(\tau) - {1 \over 2} y \tau { d \over d \tau} G(\tau) \nonumber\\
&&\tau = -y { d \over d \tau} G(\tau).
\end{eqnarray}

\n
Finally the probability distribution function can now be written as (Balian \& Schaeffer 1989):

\begin{equation}
P(\delta) = \int_{-i\infty}^{i\infty} {dy \over 2 \pi i} \exp \left [ 
{ (1+ \delta)y - \phi(y) \over \bar \xi_2} \right ]
\end{equation}

In recent studies it was found that in both the quasi-linear and the highly 
nonlinear regimes, it is possible to introduce a reduced convergence field
(Munshi \& Jain 2000, Munshi 2002),
$\eta = {\kappa - \kappa_{min} \over -\kappa_{min}}$, which to a very
good approximation follows the same statistics as $1 + \delta$. In the
quasi-linear regime it follows the smoothed projected density, and in the
highly nonlinear regime it simply follows the 3D statistics of the density 
field.\footnote{It was recently shown by Munshi (2002) that analytical 
results obtained by direct perturbative calculations can also be obtained 
by using a functional fit obtained from assuming a log-normal evolution of 
local correlated density field. This method was also found not only to work 
for one point probability distribution function but also for bias associated 
with convergence maps.}

The variance of $\eta$ is given by \cite{Valageas00a,V00b}
\begin{equation}
\label{eq:xieta}
\xi_{\eta} = \int_0^{\chi_s} {\mathrm{d}}\chi\, \left(\frac{w}{F_s}\right)^2\,
I_{\mu}(\chi),
\end{equation}
with
\ba
&& w(\chi,\chi_s) = \frac{H_0^2}{c^2}\, \frac{ {\cal{D}}(\chi)\, 
{\cal{D}}(\chi_s-\chi)} {{\cal{D}}(\chi_s)} \, (1+z), \hskip 1cm
{\cal{D}}(\chi) = \frac{cH_0^{-1}}{\sqrt{|\Omega_k|}}\,
{\rm sinn}\left(\sqrt{|\Omega_k|} \, \chi\right) \nonumber\\
&& F_s= \int _0^{\chi_s} \mathrm{d}\chi\, w(\chi, \chi_s),\hskip 1cm
I_{\mu}(z)= \pi \int_0^{\infty} \frac{\mathrm{d}k}{k}\,\,
\frac{\Delta^2(k,z)}{k}\, W^2({\cal D}k\theta_0),
\ea
where ``$\rm sinn$'' is defined as $\sinh$ if $\Omega_k>0$, and $\sin$ 
if  $\Omega_k<0$. If $\Omega_k=0$, the $\rm sinn$ and $\Omega_k$'s 
disappear. 
$\Delta^2(k,z)= 4\pi k^3 P(k,z)$, $k$ is the wavenumber,
$P(k,z)$ is the matter power spectrum, 
The window function $W({\cal D}k\theta_0)=2J_1({\cal D}k\theta_0)/
({\cal D}k\theta_0)$ for smoothing angle $\theta_0$. Here $J_1$ is the Bessel 
function of order 1.
Note that the clustering of the dark energy field would lead to an
increase in the transfer function on very large scales. Huterer (2002)
has shown that the clustering of the dark energy field can be neglected
on the scales relevant to weak lensing surveys.

Fig.1 shows $-\kappa_{min}$ and $\sqrt{\xi_{\eta}}$, and 
Figs. 2 \& 3 show the pdf for the two classes of dark energy models
listed in Table 1.

\subsection{Bias Associated with Convergence Maps} 

Assuming a correlation function hierarchy guarantees that we have a two-point
probability distribution function $P(\kappa_1,\kappa_2)$ which can be 
factorized as follows (Munshi 2001),

\begin{equation}
P(\kappa_1,\kappa_2) d\kappa_1 d\kappa_2 = P(\kappa_1)P(\kappa_2)
\left[1 + b(\kappa_1)\xi^{\kappa}_{12}b(\kappa_2)\right] d\kappa_1d\kappa_2.
\end{equation}

\n
The function $\xi^{\kappa}_{12}$ is the two point correlation function
corresponding to convergence maps.
The function $b(\kappa)$ is the bias associated with the convergence maps,
and can be shown to be related to the bias associated with overdense regions, 
$b(1+\delta)$ (Munshi 2001).
The perturbative calculations also produce similar results, although the
nature of the bias function $b(\kappa)$ changes from one regime to another.
As was the case for one point normalized moments, whose generating function
was related to the one point probability distribution function, the 
generating functions for two-point collapsed higher order correlation
functions, also known as cumulant correlators, are also related
to the bias function in a very similar manner. We write

\begin{equation}
P(\delta)b(\delta) = \int_{-i\infty}^{i\infty} {dy \over 2 \pi i}
\tau  \exp \left [ 
{ (1+ \delta)y - \phi(y) \over \bar \xi_2} \right ],
\end{equation}

\n
where $\tau$ is a generating function for the cumulant correlators.
However, it turns out that the differential bias, as we have 
written down above, is difficult to estimate from numerical data.
Therefore we work with the cumulative bias, which is the bias associated with
points where convergence maps cross a particular threshold (Munshi 2001).
Previous studies against numerical simulations showed that the bias
function describes the numerical results quite accurately. It was shown 
in earlier studies that $b(\kappa) = {b(1+\delta) \over \kappa_{min}}$.

Figures 4 \& 5 show the cumulative bias for the two classes of dark 
energy models indicated in Figure 1.
Clearly, the cumulative bias of convergence is complementary to the 
convergence pdf in probing the non-Gaussianity of gravitational clustering
and constraining dark energy models.

\subsection{A New Indicator for Deviations from the $\Lambda$CDM model}

We find that the deviations of dark energy models
from the fiducial $\Lambda$CDM model can be quantified with
a single parameter

\begin{equation}
1+ \epsilon \equiv \frac{ \sqrt{\bar{\xi_\kappa}({\mbox{XCDM}})}}
{ \sqrt{\bar{\xi_\kappa}({\Lambda\mbox{CDM}})}}
=\frac{ \kappa_{min}({\mbox{XCDM}})\, 
\sqrt{\bar{\xi_{\eta}}({\mbox{XCDM}})}  }
{ \kappa_{min}(\Lambda\mbox{CDM})\, 
\sqrt{\bar{\xi_{\eta}}(\Lambda \mbox{CDM})} },
\end{equation}
\n
where XCDM represents an arbitrary dark energy model. 
Figure 6 shows the indicator $1+\epsilon$ for the two classes of dark energy
models studied in this paper, for smoothing angle $\theta_0=1'$ and $15'$.

Comparison of Fig.6 and Figs.2-3 shows that the more the pdf
of the dark energy model differs from that of the fiducial 
$\Lambda$CDM model, the more the indicator $1+\epsilon$ deviates
from one. This indicates that the pdf is primarily determined by
its variance, which is consistent with the finding of
the existence of a universal probability distribution function (in terms of
the scaled convergence $\eta$) for weak lensing amplification 
by Wang, Holz, \& Munshi (2002).

It is useful to compare our new indicator with the convergence skewness 
$S_3$ (Hui 1999)
for the same models. Figure 7 shows $S_3$ for the same models as
in Figure 6, with the same line and arrow types.
We have computed the convergence skewness $S_3$ using Hyper-Extended 
Perturbation theory in the nonlinear regime (small smoothing angular scales), 
and perturbative results are adopted for the quasi-linear regime
(larger smoothing angular scales). While $S_3$, as an indicator, mainly
encodes the information about non-Gaussianity, the indicator we 
have proposed is directly related to the variance and is more
sensitive to the projected density power spectrum.

Our new weak lensing pdf shape indicator, $1+\epsilon$, is complementary
to $S_3$ in constraining the dark energy equation of state.
The indicator, $1+\epsilon$, is sensitive to smoothing angle $\theta_0$,
while $S_3$ is not very sensitive to $\theta_0$ at small angular scales. 
Feasible future supernova surveys
can yield a large number of type Ia supernovae out to 
redshift $z=1$ and beyond (Wang 2000, 
SNAP\footnote{See http://snap.lbl.gov.}).
It may be possible to directly measure the weak lensing pdf with sufficiently
high statistics (Metcalf \& Silk 1999; Seljak \& Holz 1999);
this would allow us to utilize the pdf with different smoothing angles
to probe different ranges of constant $w_X$ models,
and the variation of $w_X$ with $z$.

\section{Discussions and Summary} 

We have analyzed weak lensing statistics for two classes of dark energy 
cosmological models. One class of dark energy models have effective constant 
equation of state $w_X$, while the other have linear
time-varying $w_X(z)$ inspired by quintessence models. 
The weak lensing statistics of these dark energy models are compared 
with that of a ${\Lambda}$CDM model. 

It has been shown that in directly using the distance-redshift relations of
type Ia supernovae to probe dark energy, it is optimal to measure the
dark energy density, $\rho_X(z)=\rho_X(0)\,f(z)$, instead
of the dark energy equation of state, $w_X(z)$. (Wang \& Garnavich 2001;
Wang \& Lovelace 2001; Tegmark 2001)
In this paper, for convenience and illustration,
we have used $w_X$ to classify various models.

Note that we have considered a dark energy toy model which violates 
the weak energy condition, as similar models could arise from
quantum effects in quantum gravity models of inflation (Onemli \& Woodard 2002).
Also, we have only considered dark energy models with linear time-varying 
$w_X(z)$, although dark energy models with much more complicated time 
dependence in $w_X(z)$ have been proposed (see for example, 
Bassett et al. 2002).
This is because it is extremely difficult to extract the time dependence
of $w_X(z)$, even if it were a simple linear function of the redshift $z$, 
from observational data (for example, see Maor et al. 2001;
Barger \& Marfatia 2001; Wang \& Garnavich 2001; Kujat et al. 2002; 
Maor et al. 2002).

We have studied the statistics of the cosmic convergence field 
via various diagnostics including the one point probability distribution
functions and the bias associated with convergence ``hot spots''. 
The analysis was done for both the quasi-linear scales
where perturbative calculations are valid and also for very small
angular scales where hierarchical ansatz is generally used to quantify
the statistical distributions. Following earlier studies we introduce
a quantity $\kappa_{min}$ which can help us to write the observed convergence
field in terms of a reduced convergence field which in turn represents 
directly the statistics of density distribution. For large smoothing angles, 
perturbation theory predicts this quantity $\eta$ to trace the projected
density field, and on small angular scales it traces the nonlinear density
field in three dimensions. The lower order moments corresponding to various 
cosmologies have already been investigated in detail and our studies
complement these results. Also we have used top-hat window functions,
but Bernardeau \& Valageas (2000) have shown how to generalize 
similar calculations in terms of aperture mass using compensated
filters. 

We have identified a new weak lensing pdf shape indicator,
$1+\epsilon \propto \sqrt{\bar{\xi_{\kappa}}} =\kappa_{min}\, 
\sqrt{\bar{\xi_{\eta}}}$,
which can be used to predict the expected difference in weak lensing 
statistics between various dark energy models, and for choosing optimized
smoothing angles to constrain a given class of dark energy models.
For example, small smoothing angles are favored for constraining dark
energy models with $w_X < -1$.

Our proposed $1+\epsilon$ statistics is related to the volume average of 
two-point statistics of $\bar{\xi_{\kappa}}$. 
Note that while $\bar{\xi_{\eta}}$ only depends on the smoothing angle 
and the underlying mass distribution, $\kappa_{min}$ encodes
the dependence on cosmological parameters. Therefore, our new
statistical indicator, $1+\epsilon \propto \sqrt{\bar{\xi_{\kappa}}} 
=\kappa_{min}\, \sqrt{\bar{\xi_{\eta}}}$, is of interest
as far as we are interested in differentiating
various cosmological models. We found that both $P(\kappa)$ and
$b(\kappa)$ depends on $k_{min}$ and $\xi_{\eta}$, however, it is
difficult to infer the difference in these statistics in
various cosmologies. The $1+\epsilon$ statistics we have devised, 
however, is quite interesting in the sense that it can very easily
be used to check how much various dark energy models differ
in weak lensing. It can be used to complement and supplement
various other statistics such as $S_3$.

We have computed the convergence skewness $S_3$ for the dark energy models
considered in this paper. We find that the new weak lensing pdf shape 
indicator, $1+\epsilon$, is indeed complementary to $S_3$ in probing dark
energy equation of state.

We note that getting maps of convergence is difficult compared to
direct evaluation of non-Gaussian statistics from shear maps
(see e.g. Schneider \& Lombardi 2002; Zaldariaga \& Scoccimarro 2002).
On the other hand, the construction
of convergence statistics can be directly modeled at arbitrary
level, whereas for shear field the computation of statistics
is done in a order by order manner so far. So an independent analysis of
convergence maps constructed from shear maps should be useful
in constraining various errors which might get introduced during
various stages of data reduction. The question of error bars
in weak lensing measurements
has been dealt with in great detail in Munshi \& Coles (astro-ph/0003481,
to appear in MNRAS) for various window functions and are 
independent of the cosmological model assumed. 
Our convergence statistics can be a powerful diagnostic and complementary
tool to the shear map statistic.

With high statistics data from future weak lensing surveys of galaxies
and supernova pencil beam surveys (Wang 2000; SNAP), weak lensing can be
a useful tool in differentiating different dark energy models.

\acknowledgements
DM was supported by PPARC grant RG28936, and
YW was supported in part by NSF CAREER grant AST-0094335.
DM would like to thank Alexandre Refregerier for many useful discussion, and
Francis Bernardeau for making a copy of his code to compute the pdf and bias
available to us.

\clearpage
\setcounter{figure}{0}

\figcaption[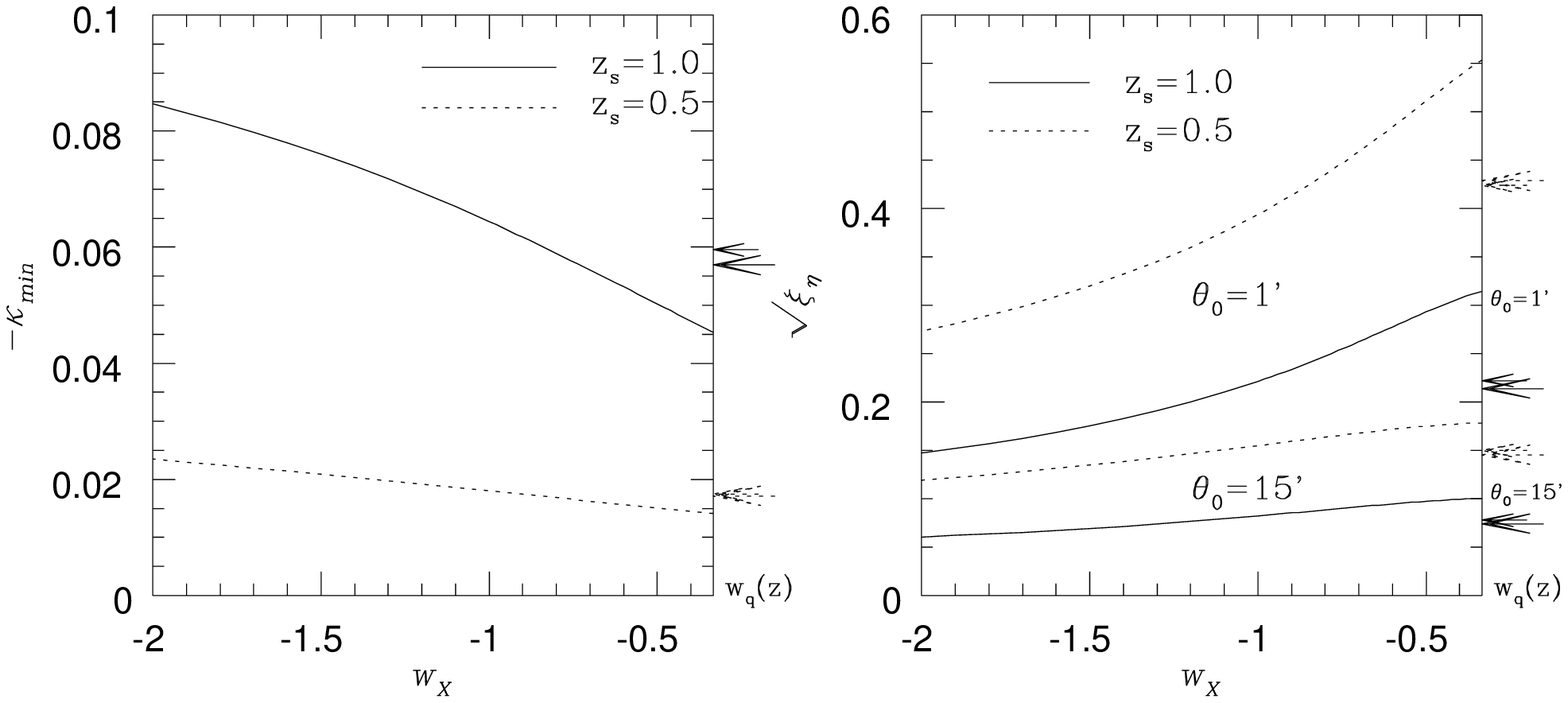]
{The minimum value of the convergence field, $\kappa_{min}$, 
 and variance of the scaled convergence $\eta=1+\kappa/|\kappa_{min}|$, 
 $\xi_{\eta}$, as functions of constant dark energy equation of state, 
 $w_X$, for source redshift $z_s=0.5$ and 1.0. 
The arrows indicate the $-\kappa_{min}$ and $\sqrt{\xi_{\eta}}$ values for 
dark energy models with time-varying equation of state, $w_q(z)=-1+z$ 
(long arrows) and $w_q(z)=-1+2z/3$ (short arrows).
Note that $\kappa_{min}$ does not depend on the smoothing angle
$\theta_0$ but it depends on the background dynamics of the universe.
}
 
\figcaption[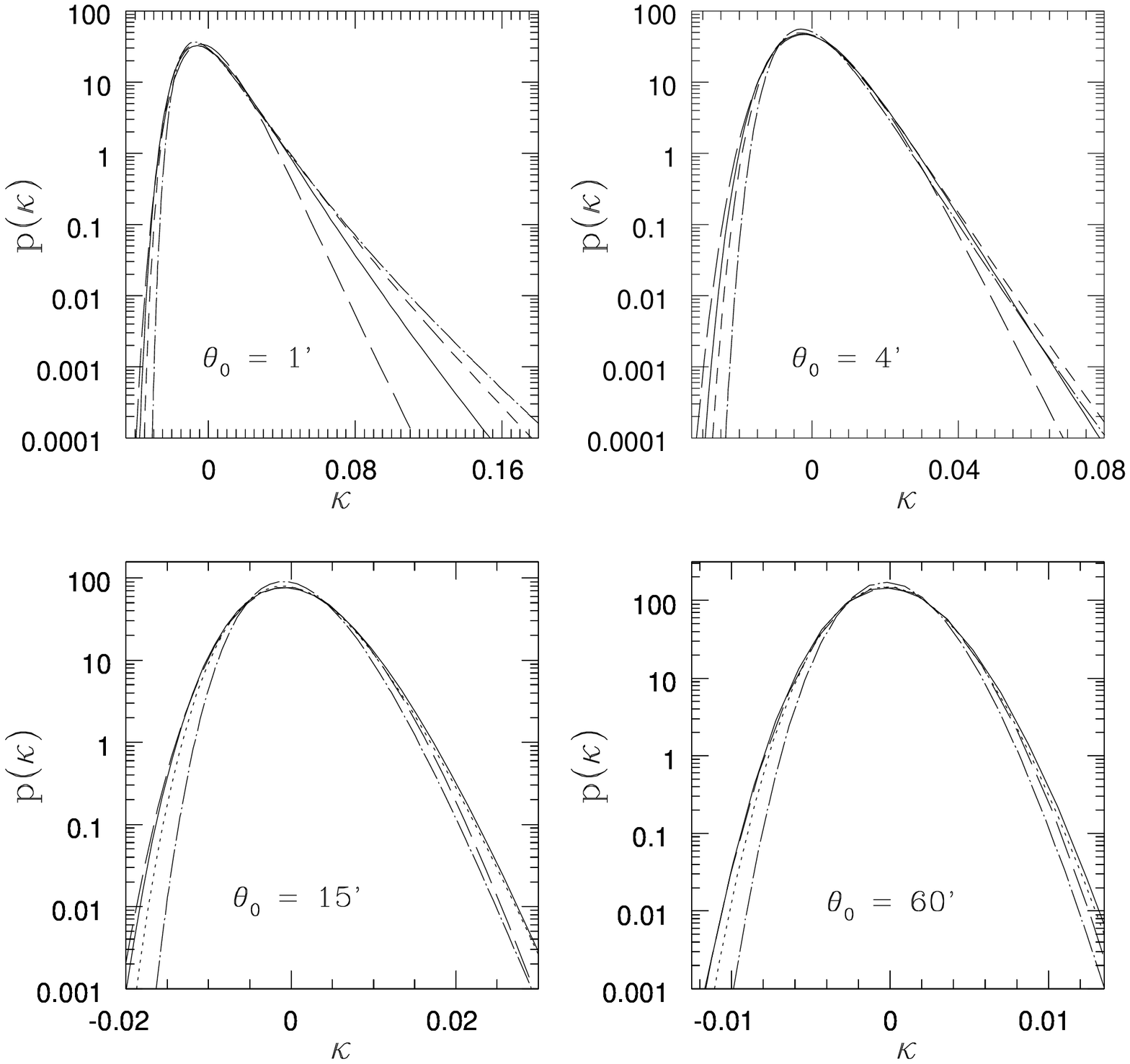]
{PDF associated with constant $w_X$ models. Two different regimes
are considered for computing the PDF. Upper panels $\theta_s = 1',4'$
 correspond to the nonlinear
calculations where we have assumed a hierarchical ansatz for the
correlation
hierarchy. For the lower panels perturbative results are used to
construct the PDF at smoothing angles $\theta_s = 15'$ and $60'$.
Various curves correspond to various quintessence models. In each panel
solid line represents the $\Lambda$CDM model, short dashed line represent
$w_X = -2/3$, dot-dashed line represents $w_X = -1/3$, and long
dashed line represents $w_X = -1.9$.
}

\figcaption[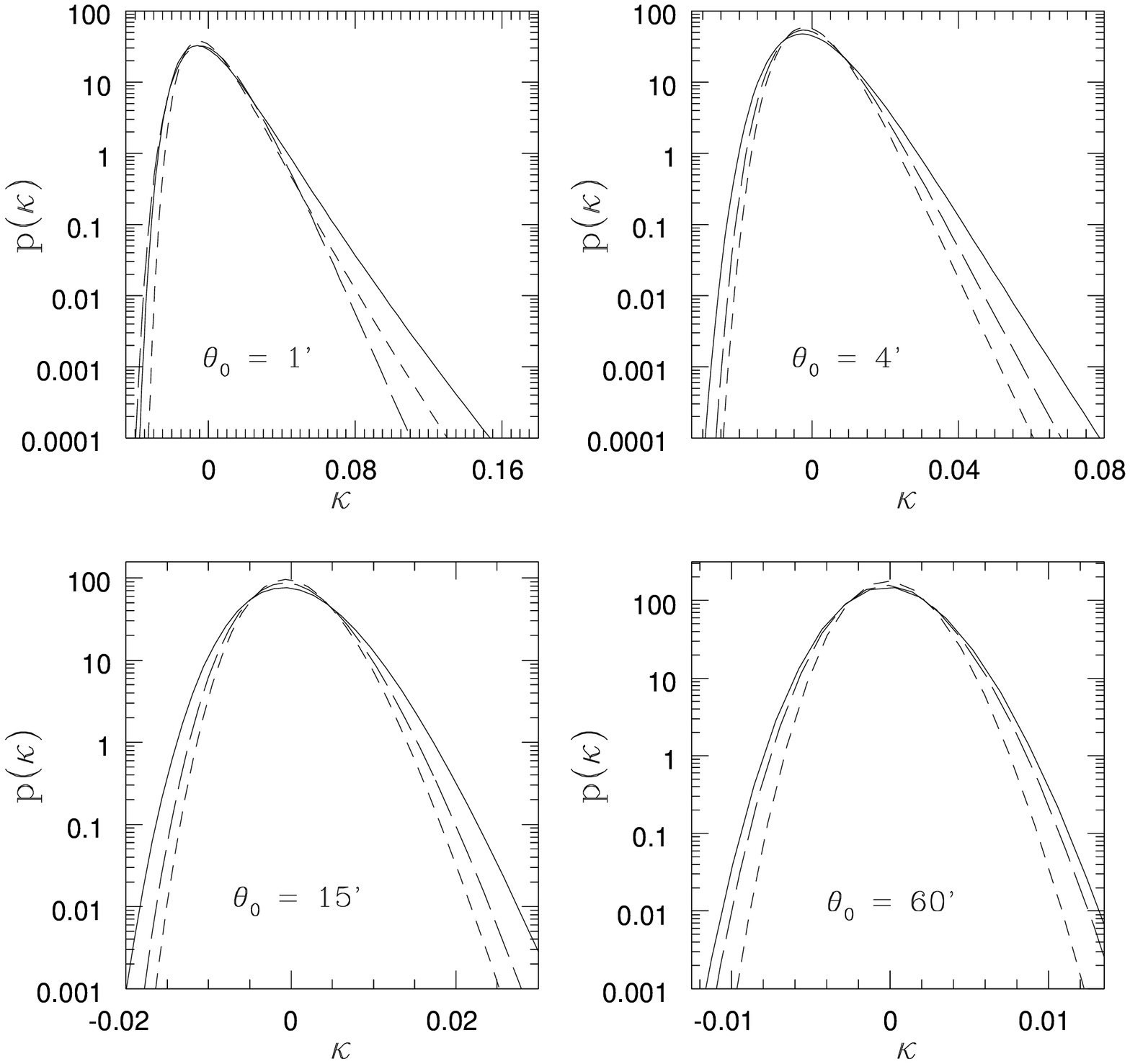]
{PDF associated with time-varying $w_X$ models compared to 
$\Lambda$CDM. As in the previous figure two different regimes
are considered for computing the PDF. Upper panels $\theta_s = 1',4'$
 correspond to the nonlinear
calculations where we have assumed a hierarchical ansatz for the
correlation
hierarchy. For the lower panels perturbative results are used to
construct the PDF at smoothing angles $\theta_s = 15'$ and $60'$.
Various curves correspond to various quintessence models. In each panel
solid line represents the $\Lambda$CDM model, short dashed line represent
$w_X = -1+z$ model, and long dashed line represents $w_X = -1+2z/3$ model.
}

\figcaption[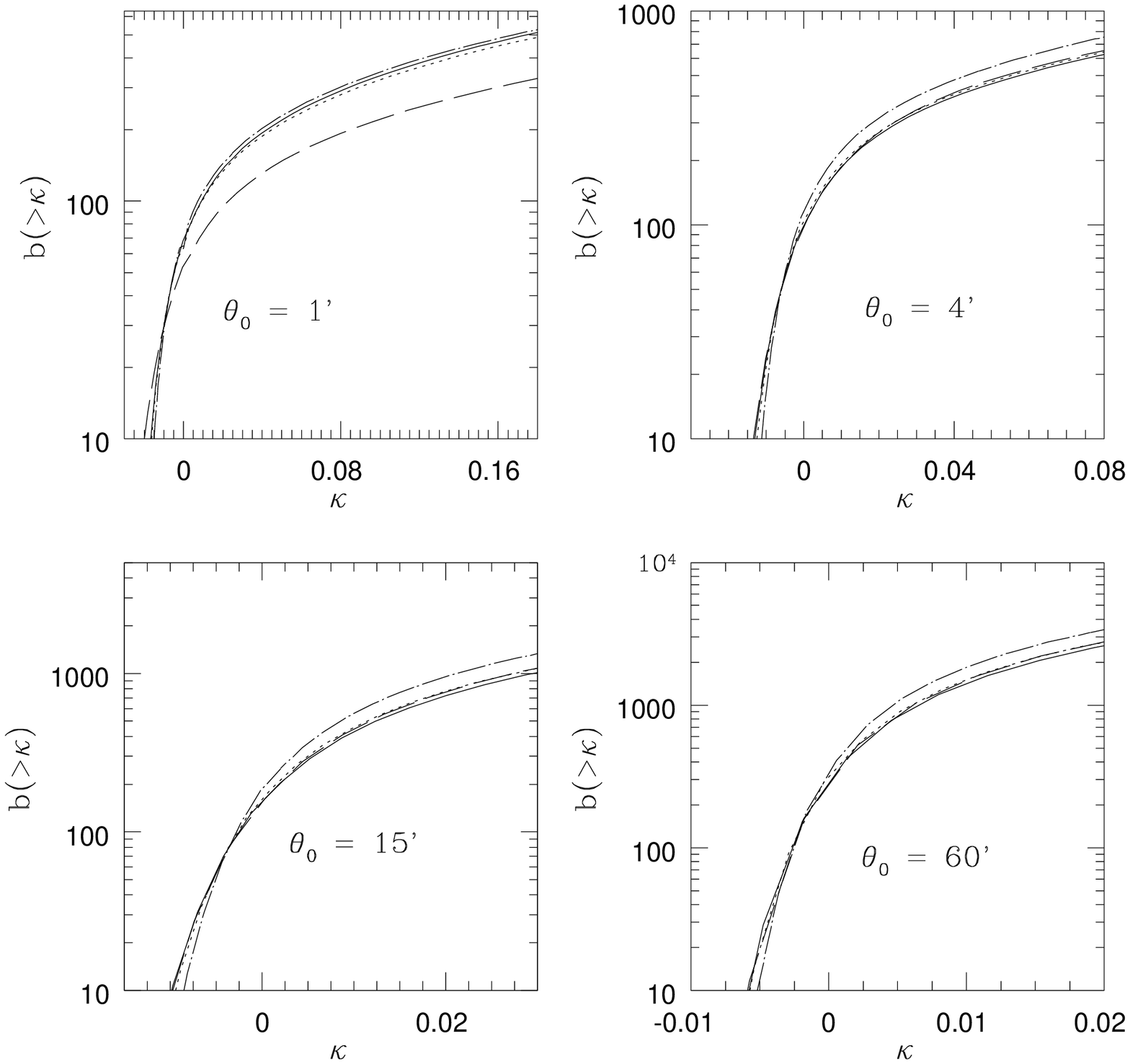]
{Bias associated with constant $w_X$ models. Two different regimes
are considered for computing the PDF. Upper panels $\theta_s = 1',4'$
correspond to the nonlinear calculations where we have assumed a hierarchical 
ansatz for the correlation
hierarchy. For the lower panels perturbative results are used to
construct the PDF at smoothing angles $\theta_s = 15'$ and $60'$.
Various curves correspond to various quintessence models. In each panel
solid line represents the $\Lambda$CDM model, short dashed line represent
$w_X = -2/3$, dot-dashed line represents $w_X = -1/3$, and long
dashed line represents $w_X = -1.9$.
}

\figcaption[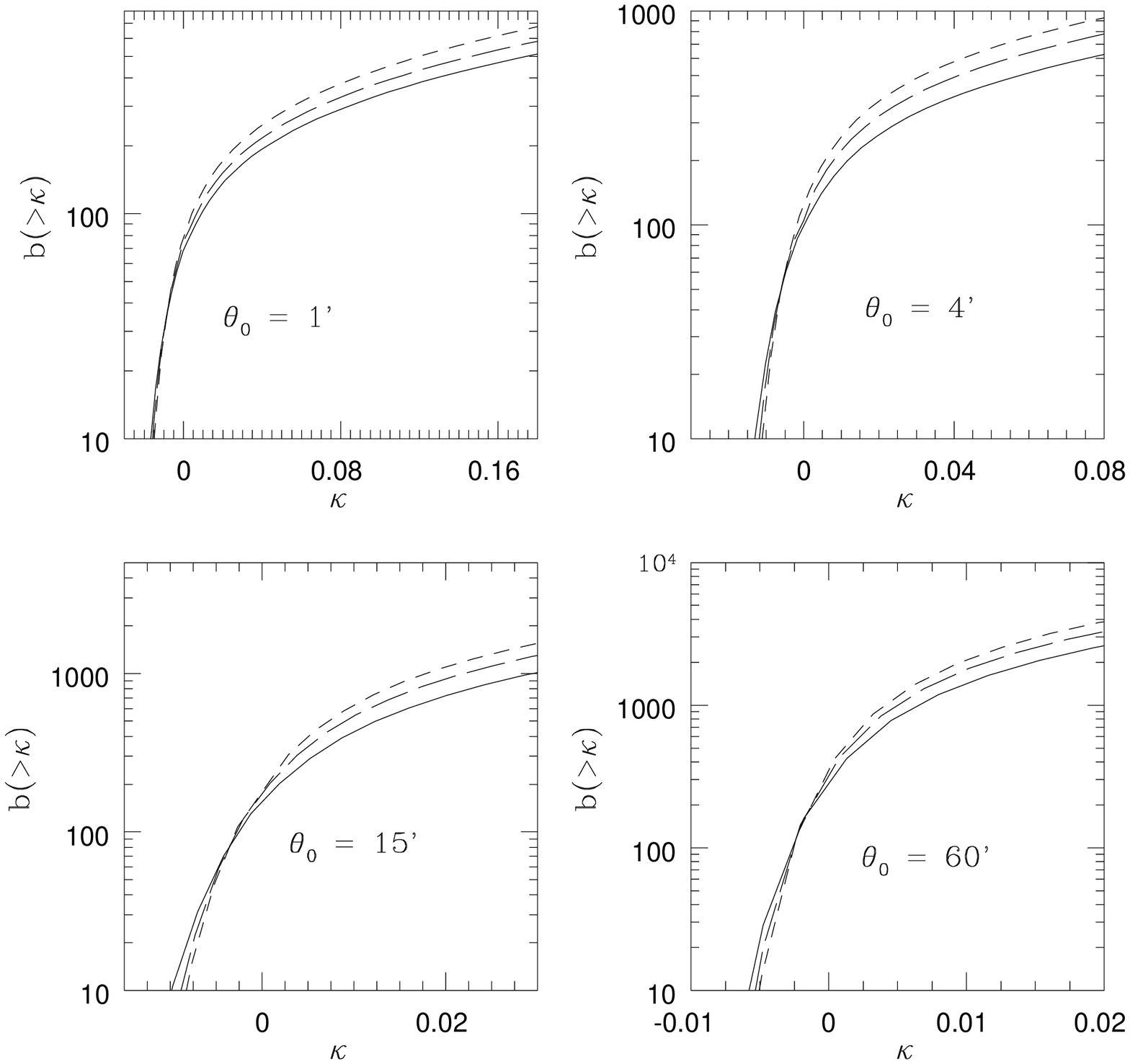]
{Bias associated with time-varying $w_X$ models compared to $\Lambda$CDM.
As in previous figure two different regimes
are considered for computing the PDF. Upper panels $\theta_s = 1',4'$
 correspond to the nonlinear
calculations where we have assumed a hierarchical ansatz for the
correlation
hierarchy. For the lower panels perturbative results are used to
construct the PDF at smoothing angles $\theta_s = 15'$ and $60'$.
Various curves correspond to various quintessence models. In each panel
solid line represents the $\Lambda$CDM model, short dashed line represent
$w_X = -1+z$ model, and long dashed line represents $w_X = -1+2z/3$ model.
}

\figcaption[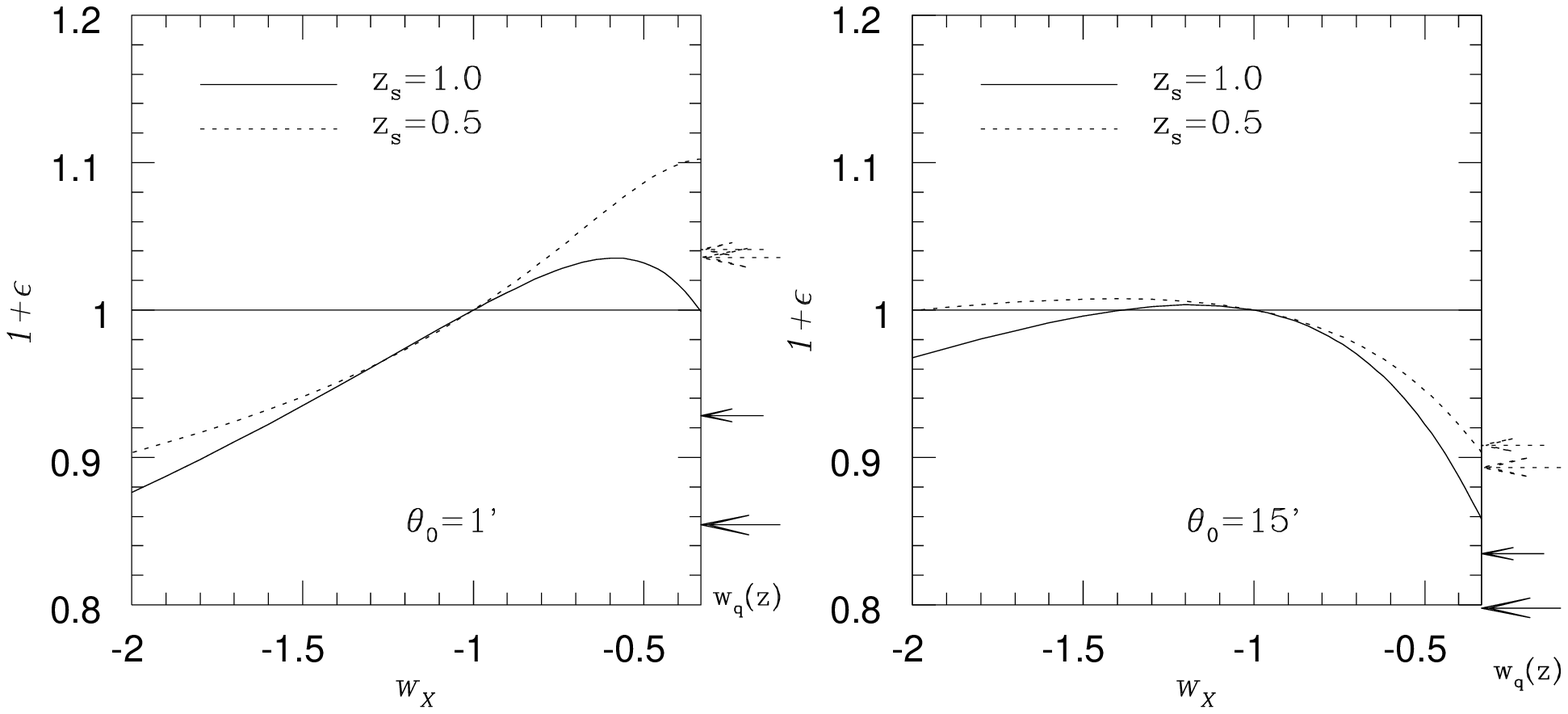]
{Indicator of the deviations of dark energy models from the fiducial 
$\Lambda$CDM model, $1+\epsilon$, as function of constant dark energy
equation of state $w_X$, for smoothing angle $\theta_0=1'$ and $15'$.
The arrows indicate the $1+\epsilon$ values for dark energy
models with time-varying equation of state, $w_q(z)=-1+z$ (long arrows),
and $w_q(z)=-1+2z/3$ (short arrows).}

\figcaption[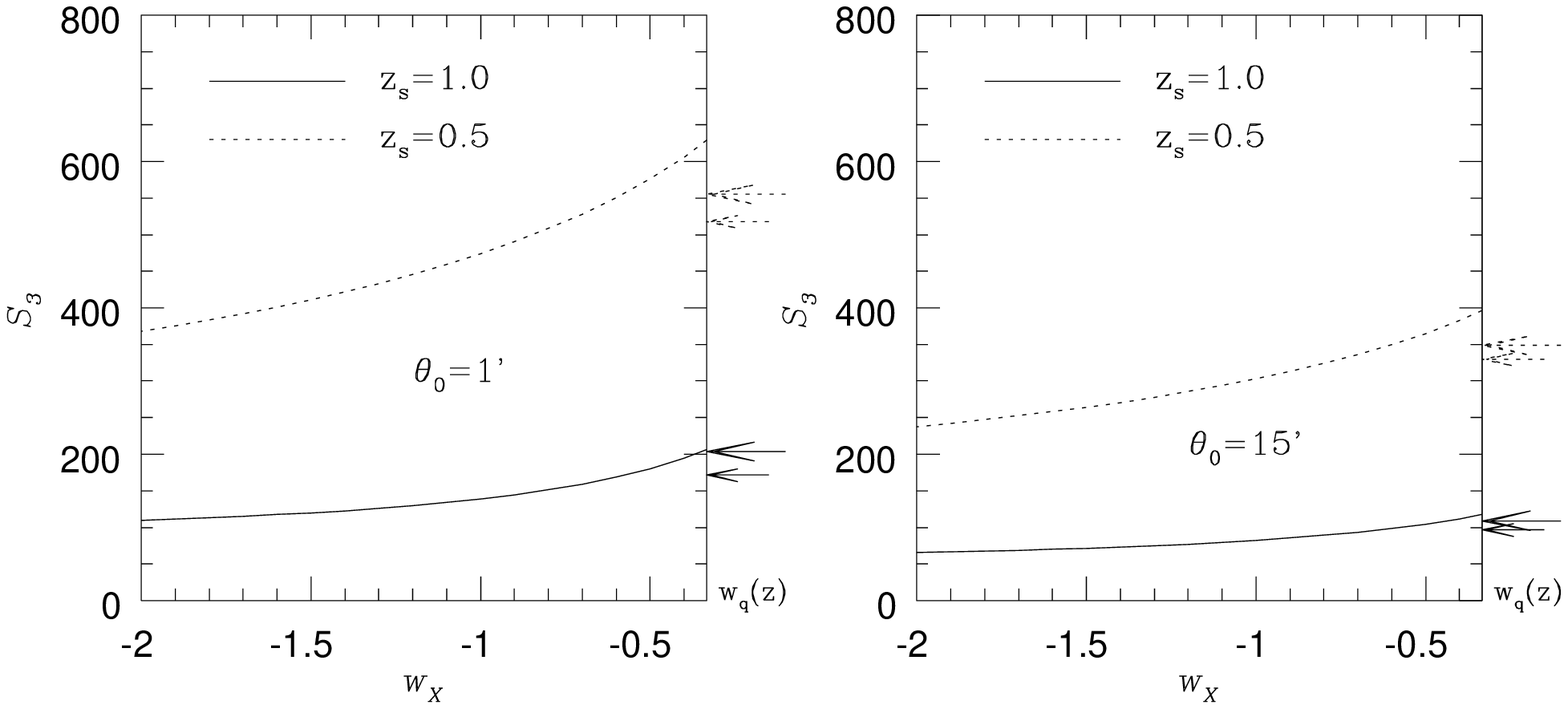]
{The skewness $S_3$, for the same models as in Fig.6.
The line and arrow types are the same. In the left panel we use Hyper-Extended 
Perturbation theory to compute the convergence skewness, whereas 
in the right panel (where larger smoothing angular scales are
considered) perturbative results are adopted.}

\clearpage
\setcounter{figure}{0}
\plotone{f1.eps}
\figcaption[f1.eps]
{The minimum value of the convergence field, $\kappa_{min}$, 
 and variance of the scaled convergence $\eta=1+\kappa/|\kappa_{min}|$, 
 $\xi_{\eta}$, as functions of constant dark energy equation of state, 
 $w_X$, for source redshift $z_s=0.5$ and 1.0. 
The arrows indicate the $-\kappa_{min}$ and $\sqrt{\xi_{\eta}}$ values for 
dark energy models with time-varying equation of state, $w_q(z)=-1+z$ 
(long arrows) and $w_q(z)=-1+2z/3$ (short arrows).
Note that $\kappa_{min}$ does not depend on the smoothing angle
$\theta_0$ but it depends on the background dynamics of the universe.
}
 
\plotone{f2.eps}
\figcaption[f2.eps]
{PDF associated with constant $w_X$ models. Two different regimes
are considered for computing the PDF. Upper panels $\theta_s = 1',4'$
 correspond to the nonlinear
calculations where we have assumed a hierarchical ansatz for the
correlation
hierarchy. For the lower panels perturbative results are used to
construct the PDF at smoothing angles $\theta_s = 15'$ and $60'$.
Various curves correspond to various quintessence models. In each panel
solid line represents the $\Lambda$CDM model, short dashed line represent
$w_X = -2/3$, dot-dashed line represents $w_X = -1/3$, and long
dashed line represents $w_X = -1.9$.
}

\plotone{f3.eps}
\figcaption[f3.eps]
{PDF associated with time-varying $w_X$ models compared to 
$\Lambda$CDM. As in the previous figure two different regimes
are considered for computing the PDF. Upper panels $\theta_s = 1',4'$
 correspond to the nonlinear
calculations where we have assumed a hierarchical ansatz for the
correlation
hierarchy. For the lower panels perturbative results are used to
construct the PDF at smoothing angles $\theta_s = 15'$ and $60'$.
Various curves correspond to various quintessence models. In each panel
solid line represents the $\Lambda$CDM model, short dashed line represent
$w_X = -1+z$ model, and long dashed line represents $w_X = -1+2z/3$ model.
}

\plotone{f4.eps}
\figcaption[f4.eps]
{Bias associated with constant $w_X$ models. Two different regimes
are considered for computing the PDF. Upper panels $\theta_s = 1',4'$
correspond to the nonlinear calculations where we have assumed a hierarchical 
ansatz for the correlation
hierarchy. For the lower panels perturbative results are used to
construct the PDF at smoothing angles $\theta_s = 15'$ and $60'$.
Various curves correspond to various quintessence models. In each panel
solid line represents the $\Lambda$CDM model, short dashed line represent
$w_X = -2/3$, dot-dashed line represents $w_X = -1/3$, and long
dashed line represents $w_X = -1.9$.
}

\plotone{f5.eps}
\figcaption[f5.eps]
{Bias associated with time-varying $w_X$ models compared to $\Lambda$CDM.
As in previous figure two different regimes
are considered for computing the PDF. Upper panels $\theta_s = 1',4'$
 correspond to the nonlinear
calculations where we have assumed a hierarchical ansatz for the
correlation
hierarchy. For the lower panels perturbative results are used to
construct the PDF at smoothing angles $\theta_s = 15'$ and $60'$.
Various curves correspond to various quintessence models. In each panel
solid line represents the $\Lambda$CDM model, short dashed line represent
$w_X = -1+z$ model, and long dashed line represents $w_X = -1+2z/3$ model.
}

\plotone{f6.eps}
\figcaption[f6.eps]
{Indicator of the deviations of dark energy models from the fiducial 
$\Lambda$CDM model, $1+\epsilon$, as function of constant dark energy
equation of state $w_X$, for smoothing angle $\theta_0=1'$ and $15'$.
The arrows indicate the $1+\epsilon$ values for dark energy
models with time-varying equation of state, $w_q(z)=-1+z$ (long arrows),
and $w_q(z)=-1+2z/3$ (short arrows).}

\plotone{f7.eps}
\figcaption[f7.eps]
{The skewness $S_3$, for the same models as in Fig.6.
The line and arrow types are the same. In the left panel we use Hyper-Extended 
Perturbation theory to compute the convergence skewness, whereas 
in the right panel (where larger smoothing angular scales are
considered) perturbative results are adopted.}

\end{document}